\documentclass[a4paper,fleqn,usenatbib]{mnras}

\newcommand{\chandra}{{\it Chandra}}
\newcommand{\xmm}{{\it XMM-Newton}}
\newcommand{\rxte}{{\it RXTE}}
\newcommand{\lx}{erg~s$^{-1}$}

\usepackage{newtxtext,newtxmath}
\usepackage{mathtools}

\usepackage[T1]{fontenc}
\usepackage{ae,aecompl}
\usepackage{romannum}

\usepackage{graphicx}	
\usepackage{amsmath}	
\usepackage{amssymb}	

\title[Pulsed fraction vs luminosity]{Anti-correlation between X-ray luminosity and pulsed fraction in the Small Magellanic Cloud pulsar SXP 1323}
\author[J. Yang et al.]{\parbox{\linewidth}{\centering Jun Yang,$^{1}$\thanks{E-mail: junyang@astro.utah.edu} Andreas Zezas,$^{2}$ Malcolm J. Coe,$^{3}$ Jeremy J. Drake,$^{2}$\\ JaeSub Hong,$^{2}$ 
Silas G. T. Laycock,$^{4}$ 
and Daniel R. Wik$^{1}$ %
}%
\\
$^{1}$Department of Physics and Astronomy, the University of Utah\\
$^{2}$Harvard-Smithsonian Center for Astrophysics, Cambridge, MA 02138, USA\\
$^{3}$Physics \& Astronomy, University of Southampton, SO17 1BJ, UK\\
$^{4}$Department of Physics and Applied Physics, University of Massachusetts, Lowell, MA 01854, USA
}

\date{Accepted 2018 May 8. Received 2018 May 8; in original form 2018 January 3}

\pubyear{2017}

\begin{document}
\label{firstpage}
\pagerange{\pageref{firstpage}--\pageref{lastpage}}
\maketitle

\begin{abstract}
We report the evidence for the anti-correlation between pulsed fraction (PF) and luminosity of the X-ray pulsar SXP 1323, found for the first time in a luminosity range $10^{35}$--$10^{37}$ erg s$^{-1}$ from observations spanning 15 years. The phenomenon of a decrease in X-ray PF when the source flux increases has been observed in our pipeline analysis of other X-ray pulsars in the Small Magellanic Cloud (SMC). It is expected that the luminosity under a certain value decreases as the PF decreases due to the propeller effect. Above the propeller region, an anti-correlation between the PF and flux might occur either as a result of an increase in the un-pulsed component of the total emission or a decrease of the pulsed component. Additional modes of accretion may also be possible, such as spherical accretion and a change in emission geometry. At higher mass accretion rates, the accretion disk could also extend closer to the neutron star (NS) surface, where a reduced inner radius leads to hotter inner disk emission. These modes of plasma accretion may affect the change in the beam configuration to fan-beam dominant emission. 
\end{abstract}


\section{Introduction}
X-ray pulsars comprise two stars, a NS (descended from a star with initial mass $>$ 8 M$_\odot$, \citealt{ver95}), and a mass-losing companion star, also of large mass. The general picture of accretion onto X-ray pulsars consists of a flow in a wind or disk to the magnetosphere and then along the dipole field lines onto the magnetic poles of the NS. 
The pulsed fraction (PF), i.e., the relative amplitude of the emerging pulse profile, bears key information on the relation between X-ray emission from the accretion column (pulsed emission) and other regions of the accretion flow or NS surface (un-pulsed emission), e.g., \citet{bel02}. 


The SMC pulsar SXP 5.05 was reported by \citet{coe15} to show a positive correlation between the PF and luminosity, as shown in their Fig. 11. Those data were taken while SXP 5.05 was undergoing high levels of accretion. At low mass accretion rates, \citet{cui97} reported two X-ray pulsars (GX 1+4 and GRO J1744-28) whose PF decreases as the X-ray flux drops below a certain threshold. This is an indication of the propeller effect \citep{Illarionov and Sunyaev 1975} that takes place when the pulsar magnetosphere grows beyond the co-rotation radius, and the centrifugal force prevents accreting matter from reaching the magnetic poles.



\citet{tie05} observed an anti-correlation between PF and the corresponding flux of 1E 1048.1-5937 in the Milky Way. Spectral variations as a function of the pulse phase shows the hardest spectrum at pulse maximum. 
\citet{lut09} presented marginal evidence for an anti-correlation of PF and energy in source 4U 0115+63 and Her X-1. Fig. 4 from \citet{tsy07} shows the increase of energy in 4U 0115+63 is not uniform but has local maximum near the cyclotron line. A positive and an anti- correlation is observed at low and high energy, respectively. 

\citet{tsy10} noted the PF of V 0332+53 increases with decreasing photon energy below 12--15 keV, which is difficult to explain. An anti- and a positive correlation is observed at low and high luminosities, respectively (see their Fig. 10). Below $\sim$$10^{38}$ {\lx} the anti-correlation is in accordance with a geometry model in which the PF is determined by the luminosity-dependent visible areas of the accretion columns. However, In the photon energy range 25--45 keV the observed correlation does not fully conform to the model.
\citet{par89} applied a geometric model to describe the pulse shape of X-ray pulsar EXO 2030+375 and showed 
that below a luminosity of 4 $\times$ $10^{36}$ {\lx}, the pencil beam becomes dominant compared to the fan-beam, along with an increase in the un-pulsed component and a decrease in the luminosity. In \citet{bel02}'s classes of pulse profiles, visibility of the two polar caps depends on the angle between the magnetic rotation axis and dipole axis. If both poles are continuously visible, it is possible to have no pulsations. As shown in the modeled light curves of Fig. 5 from \citet{yan17b}, in classes 2 and 4 of the upper panel,
and classes 3 and 4 of the middle panel, when
both hot spots are visible, the observed pulse shows a plateau.

We have collected and analyzed the comprehensive archive of {\itshape XMM-Newton\/} (116), {\itshape Chandra\/} Advanced CCD Imaging Spectrometer (ACIS-I) (151), and {\itshape RXTE\/} (952) observations of the known pulsars in the Small Magellanic Cloud (SMC), spanning the years 1997-2014. Our pipeline generates a suite of products for each pulsar detection: spin period, flux, event list, high time-resolution light-curve, pulse-profile, periodogram, and spectrum. Combining all three satellites, we generated complete histories of the spin periods, pulse amplitudes, pulsed fractions and X-ray luminosities \citep{yan17}. 
Based on this archive, the relationship between the pulsed fraction (PF) and luminosity of the Small Magellanic Cloud pulsars have drawn our attention.
We find a surprising anti-correlation between pulsed fraction (PF) and luminosity in SMC X-ray pulsars, for example, SXP 1323, SXP 893, SXP 756, SXP 726, SXP 701, SXP 348, SXP 323. In this work, we show an example (SXP 1323) of this phenomenon and discuss the mechanism behind these results. We selected this source because it has the most data points compared to the other pulsars with anti-correlations.

SXP 1323 (a.k.a. RX J0103.6-7201) was discovered by \citet{hab05} and shows one of the longest pulse periods known in the SMC. The names of the optical companion star are [MA93] 1393 \citep{mey93} and [M2002] SMC 56901 \citep{2002ApJS..141...81M}. \citet{car17} found the orbital period of this Be/X-ray binary (BeXB) to be 26.2 days, which is very short for such a long spin period pulsar. It is located at RA = 01:03:37.5 and Dec. = -72:01:33 with a positional uncertainty of 1.1 arcsec \citep{lin12}. The spectral type of this X-ray binary counterpart is B0 with a luminosity class of \Romannum{3}-\Romannum{5} \citep{mcb08}. 

In this paper, we present the pulsed fraction dependence on luminosity from 15-years of X-ray monitoring for SXP 1323. We aim to have a deeper understanding of the accretion process under the anti-correlation of the PFs and luminosities.




\section{Observations and Methods}

\label{sec:obs} 
\citet{yan17} have collected and analyzed 36 {\xmm} and 108 {\chandra} X-ray observations up until 2014 for SXP 1323. {\xmm} has detected this source 36 times and in 10 of these observations its pulsations are found. As for {\chandra}, we only used the ACIS-I detections: 63 out of 108 observations yield source detections and 14 observations have detected its pulsations. We are not including {\rxte} observations in this analysis since RXTE does not provide the required PF information. The {\rxte} Proportional Counter Array is a non-imaging detector and multiple sources are always in the field of view, so the un-pulsed component cannot be reliably measured. 

The observations we used for SXP 1323 with pulsations detected are shown in Table~\ref{tab:obsxmm}. The pulsations are with a significance of $s \ge 99\%$ according to equation (2) of \citet{yan17}. 

In order to test the correlation with luminosities and make the results convincible, 3 different definitions of PF were calculated by integrating over the pulse
profile. The simplified formulas are shown in equations~(\ref{eq:pfa})-(\ref{eq:pfs}).

\begin{figure}
	\includegraphics[width=0.397\textwidth]{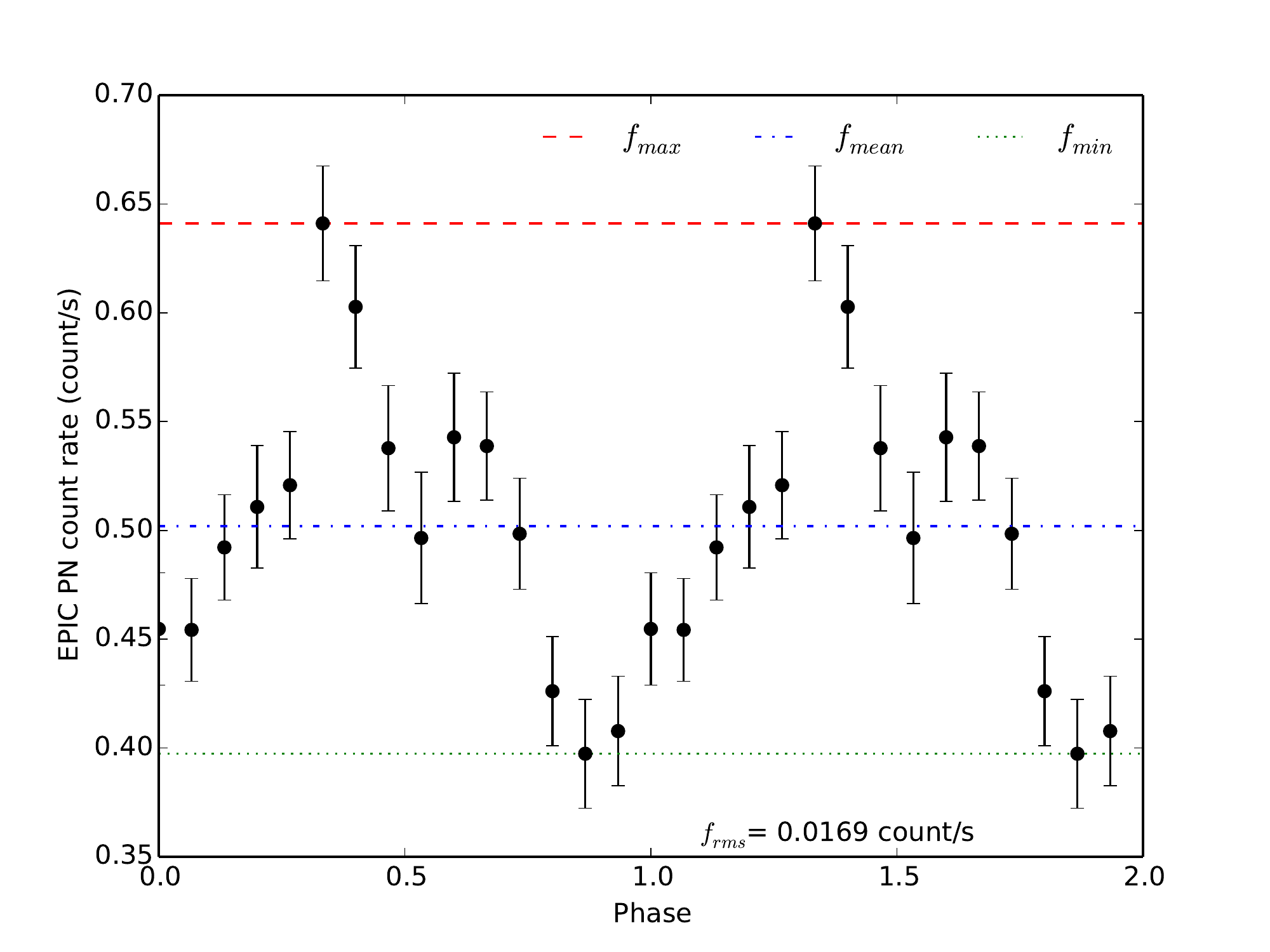}
    \caption{An example of pulse profile for SXP 1323 shows the values used for the PF calculation in equations~(\ref{eq:pfa})-(\ref{eq:pfs}); and $f_\mathrm{rms}$ is the root mean square flux. It is an {\xmm} Observation (ID 135722701), observed on 2006-04-20.}
    \label{fig:pf-cal}
\end{figure}

\begin{table}
	\centering
	\caption{The {\xmm} and {\chandra} ACIS-I X-ray observations in which the pulsations for SXP 1323 have been detected. The first column is the observation ID, the second and third columns show observing Modified Julian Date (MJD) and source flux, and the last two columns are the photon counts (for {xmm}, it is the medium value from the 3 EPIC instruments) and exposure time.}
	\label{tab:obsxmm}
	\begin{tabular}{lcccr} 
		\hline
		{\xmm}  & MJD & flux &photon  & exposure \\
		    ID          &          &(erg~s$^{-1}$~cm$^{-2})$&counts&time (ks)\\
		
		\hline
135722401&53292.38 &7.32$\times 10^{-13}$&458&31.11\\ 
123110301&51651.15&1.50$\times 10^{-12}$&2020&21.66\\ 
135722701&53845.10&4.29$\times 10^{-12}$&5096&30.48\\ 
135720801&52268.75 &2.10$\times 10^{-12}$&2190&35.02\\ 
135721701&52959.26&1.41$\times 10^{-12}$&3752&27.36\\\ 
135722501&53477.93 &4.69$\times 10^{-12}$&9129&37.12\\ 
412980201&54215.52&2.50$\times 10^{-12}$&2407&36.42\\ 
135721901&53123.30&3.21$\times 10^{-13}$&1251&35.23\\ 
412980501&54575.39&3.44$\times 10^{-12}$&3383&29.92\\ 
412980301&54399.41&1.80$\times 10^{-12}$&2725&37.12\\ 
		\hline
		
{\chandra} ID &-& - & - & - \\
		\hline
1533&52065.27& 2.33$\times 10^{-12}$&984 & 7.42\\
1536&52065.57 & 2.04$\times 10^{-12}$&860 & 7.42\\
1542&52065.76& 1.69$\times 10^{-12}$&699& 7.42\\
1786&51728.55&1.66$\times 10^{-12}$&738&7.58\\
2841&52249.09 &1.62$\times 10^{-12}$&686&7.46\\
6050&53352.15 &8.25$\times 10^{-13}$&336&7.16\\
6052&53353.37 &8.96$\times 10^{-13}$&358&7.54\\
6056&53356.31 &5.064$\times 10^{-13}$&253&8.01\\
6060&53534.55 & 9.00$\times 10^{-13}$&1033&19.8\\
6749& 53816.60&1.61$\times 10^{-12}$&1830&19.51\\
6757&53891.55& 5.24$\times 10^{-13}$&582&19.8\\
8361& 54136.10& 1.14$\times 10^{-12}$&1283&19.79\\
8364&54142.57&4.71$\times 10^{-13}$&253&8.45\\
9693&54501.17&1.55$\times 10^{-12}$&657&7.68\\
		\hline		
		
	\end{tabular}
\end{table}




\begin{equation}
    PF_\mathrm{A}=\frac{f_\mathrm{max}-f_\mathrm{min}}{f_\mathrm{max}},
	\label{eq:pfa}
\end{equation} 
here $f_\mathrm{max}$ is the maximum photon count rate in the pulse profile and $f_\mathrm{min}$ is the minimum value as demonstrated in an example of the pulsed profile in Fig.~\ref{fig:pf-cal}. $PF_\mathrm{A}$ is also usually called modulation amplitude. 

\begin{equation}
    PF_\mathrm{B}=\frac{f_\mathrm{mean}-f_\mathrm{min}}{f_\mathrm{mean}},
	\label{eq:pfb}
\end{equation} 
$f_{mean}$ is the average flux.
\begin{equation}
    PF_\mathrm{S}=\frac{\sqrt{2}f_\mathrm{rms}}{f_\mathrm{mean}},  ~  (and ~ f_\mathrm{rms}=\frac{\sqrt{\sum_i^{N}(f_i-f_\mathrm{mean})^2}}{N})
	\label{eq:pfs}
\end{equation} 
where $f_\mathrm{rms}$ is the root mean square (rms) flux, N is the number of bins for each folded light curve,
and $f_i$ is the mean photon count rate in each bin.
For a sinusoid wave, which is a good approximation to most accretion pulsars, the peak-to-peak pulsed flux $f_\mathrm{pulsed} = f_\mathrm{mean}-f_\mathrm{min}$ = $\sqrt{2}f_\mathrm{rms}$; for a square wave $f_\mathrm{pulsed} = f_\mathrm{rms}$ \citep{bil97}.

The error of the PF is calculated as following. Firstly get the error of the flux in each bin of the light curve,
\begin{equation}
    error_i=\frac{\sqrt{\sum_j^{n}(f_i-F_j)^2}}{n}, 
	\label{eq:erri}
\end{equation} 
where $F_j$ is the flux in the $i$th bin. n is the number of points in each bin.
Then we could get the error of $f_\mathrm{max}$ ($error_\mathrm{max}$) as well as the error from $f_\mathrm{min}$ ($error_\mathrm{min}$). The error of $PF_\mathrm{A}$ is:
\begin{equation}
    error_\mathrm{PFA}=\sqrt{\frac{error_\mathrm{max}^2+error_\mathrm{min}^2}{(f_\mathrm{max}-f_\mathrm{min})^2}+(\frac{error_\mathrm{max}}{f_\mathrm{max}})^2}\ast PF_\mathrm{A}
	\label{eq:erra}
\end{equation} 

In order to calculate $error_\mathrm{PFB}$ and $error_\mathrm{PFS}$, first calculate the error of the pulsed flux:
\begin{equation}
    e_\mathrm{pulse}=\frac{\sum_i^{N}\sqrt{error_\mathrm{max}^2+error_i^2}}{N},
	\label{eq:epulse}
\end{equation} 
\begin{equation}
    error_\mathrm{PFB}=\frac{e_\mathrm{pulse}}{f_\mathrm{mean}};
	\label{eq:errb}
\end{equation} 
\begin{equation}
    error_\mathrm{PFS}=\sqrt{2}\frac{e_\mathrm{pulse}}{f_\mathrm{mean}}\ast PF_\mathrm{S}.
	\label{eq:errs}
\end{equation} 

Note, in \citet{yan17} the pulsed fraction from {\xmm} is $PF_\mathrm{B}$ and the ones from {\chandra} are $PF_\mathrm{A}$. Here we used $PF_\mathrm{A}$ and $PF_\mathrm{B}$ for both {\xmm} and {\chandra} observations.


$PF_\mathrm{A}$ has intuitive appeal, but it is more difficult to determine the $f_\mathrm{min}$ than $f_\mathrm{mean}$ \citep{bil97}. People generally use $PF_\mathrm{A}$ for light curves from long time exposures, where signal-to-noise ratio is large. $PF_\mathrm{B}$ is smaller than $PF_\mathrm{A}$, but more stable as $f_\mathrm{mean}$ is easier to be determined than $f_\mathrm{max}$. $PF_\mathrm{S}$ is used for relatively short time exposure. 



\begin{figure}
	\includegraphics[width=0.473\textwidth]{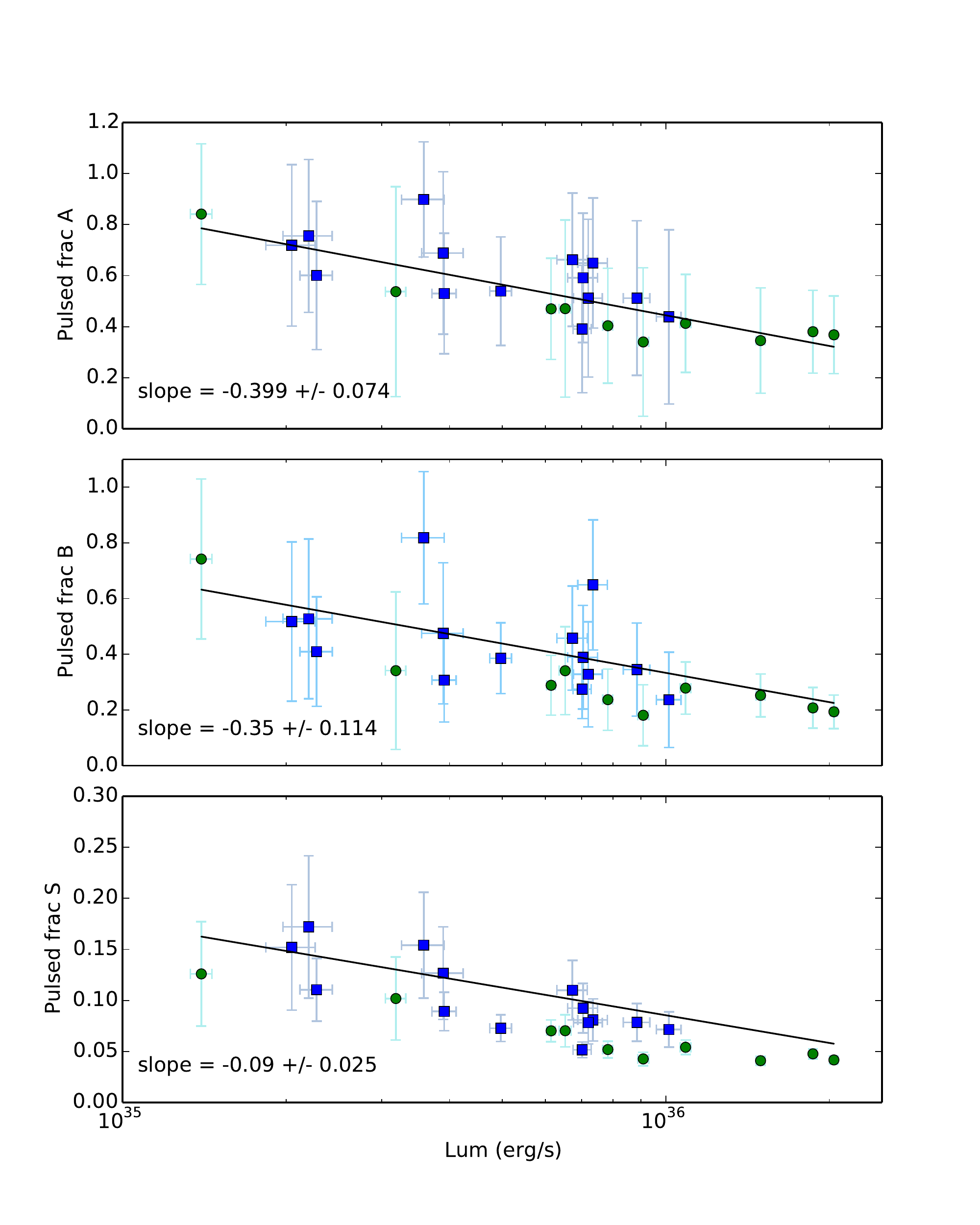} 
    \caption{The PF as the function of luminosity for SXP 1323. Green circles are the {\xmm} Detections and Blue square symbols present {\chandra} observations. The 3 panels show the PFs with different calculations which are in equations~(\ref{eq:pfa})-(\ref{eq:pfs}).}
    \label{fig:pf-lum}
\end{figure}

The PF as a function of luminosity for SXP 1323 is shown in Fig.~\ref{fig:pf-lum}. Although the light-curves were extracted from the higher time-resolution EPIC-PN data \citep{yan17}, the luminosities used in Fig.~\ref{fig:pf-lum} were obtained from the total {\xmm} flux available in the 3 {\xmm} catalogue since they are more complete than the instrument-specific fluxes. These fluxes are based on a spectral model of a power-law of slope 1.7 absorbed by a Hydrogen column of $3 \times 10^{20}$ cm$^{-2}$ (0.2-12 keV) \footnote{\url{https://www.cosmos.esa.int/web/xmm-newton/uls-userguide}}.


The trend between $PF_\mathrm{A}$ and luminosity is:
\begin{equation}
    PF_\mathrm{A}=-0.399~log (\frac{L_\mathrm{X}}{10^{35} ~\text{erg/s}})+0.850,  
	\label{eq:fit-a}
\end{equation} 

The fit between $PF_\mathrm{B}$ and luminosity is:
\begin{equation}
    PF_\mathrm{B}=-0.350~log (\frac{L_\mathrm{X}}{10^{35} ~\text{erg/s}})+0.669, 
	\label{eq:fit-b}
\end{equation}

The anti-correlation of $PF_\mathrm{S}$ and luminosity is fitted as:
\begin{equation}
    PF_\mathrm{S}=-0.101~log (\frac{L_\mathrm{X}}{10^{35} ~\text{erg/s}})+0.173, 
	\label{eq:fit-s}
\end{equation} 
where $L_\mathrm{X}$ is in erg/s.

The trend with $PF_\mathrm{A}$ is steeper than the one with $PF_\mathrm{B}$, and even more steeper than $PF_\mathrm{S}$. $PF_\mathrm{A}$ is the most popular way to show the pulsed fraction of the X-ray pulse profiles, therefore, the linear regression is more convincing. However, all of them show the similar anti-correlation.


Monte Carlo simulations are performed to estimate the false-positive detection rate for the correlation between these two observables, from which the significance level is estimated.

\begin{figure}
	\includegraphics[width=0.45\textwidth]{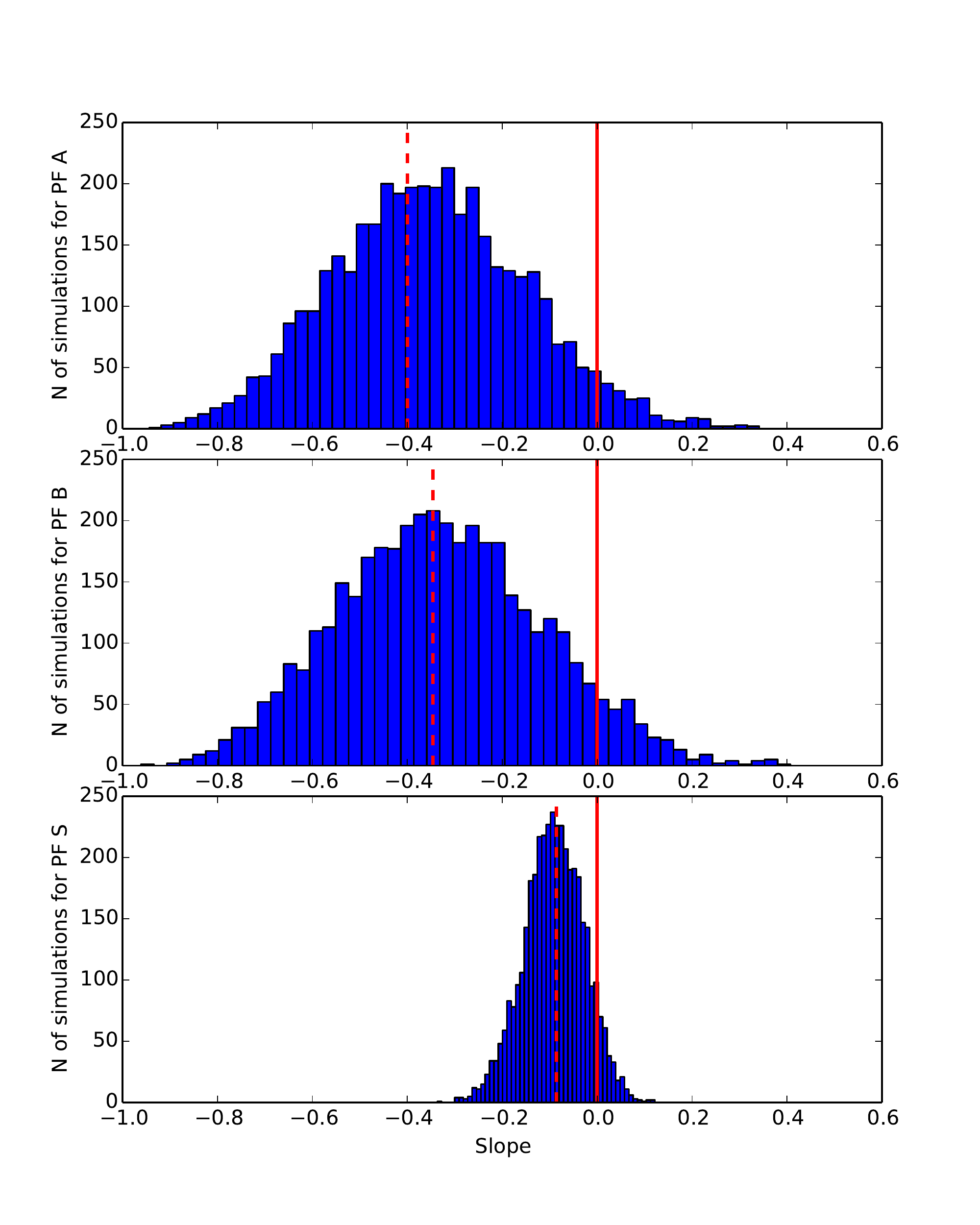} 
    \caption{Frequency distribution of correlation slopes for $PF_\mathrm{A}$ (upper), $PF_\mathrm{B}$ (middle) and $PF_\mathrm{S}$ (bottom) obtained using Monte Carlo method with 4000 simulations. The heights of bars indicate the number of parameter values in the equally spaced bins. The limit for false positive detections of an
anti-correlation is shown as red solid lines at slope 0.0. The dashed lines are the slopes from Fig.~\ref{fig:pf-lum}.}
    \label{fig:sig}
\end{figure}

For the correlation in each panel of Fig.~\ref{fig:pf-lum}, 4000 trial generates 4000 simulated data. Based on these data, the fitting is performed. One of the fitting parameters (slope) is shown in the histograms of Fig.~\ref{fig:sig}. We interpret positive slopes as false positive detections of an
anti-correlation in the real data. The number of false positives from
Fig.~\ref{fig:sig} correspond to a probability of 95.43, 93.28 and, 92.68, for the
anti-correlation found by using $PF_\mathrm{A}$, $PF_\mathrm{B}$, and $PF_\mathrm{S}$,
respectively.
Therefore, the fit of the correlations in Fig.~\ref{fig:pf-lum} is around $\sim$2 $\sigma$ confidence. 


\section{Theoretical mechanisms}

\citet{muk00} observed a decrease in the pulsed fraction with decreasing luminosity of the X-ray pulsar Cepheus X-4 (GS 2138+56). 
 However, they argued that the decrease in the pulsed fraction, depending on the accretion flow geometry with respect to line of sight, is not a consequence of propeller effect. They propose as a more likely scenario a different mode of accretion occurring below a certain luminosity. These additional entry modes of plasma may affect the emission geometry to be more fan-beam like pattern, which will increase the un-pulsed flux, and the pulsed fractions end up being smaller.

However, for SXP 1323, the PF increases as the luminosity decreases. The critical luminosity mentioned in \citet{muk00} is the maximum luminosity $L_\mathrm{X} (\mathrm{min})$ at which the centrifugal inhibition dominates, resulting in the propeller effect \citep[e.g.,][]{sht05, tsy16}: 
  \begin{equation}
  \begin{split}
    L_\mathrm{X} (\mathrm{min}) = & 2 \times 10^{37}(\frac{R}{10^6 cm})^{-1}(\frac{M}{1.4M_{\odot}})^{-\frac{2}{3}} \\
             &\times(\frac{\mu}{10^{30}~ G~ cm^{3}})^{2}(\frac{P_\mathrm{s}}{1 s})^{-\frac{7}{3}} erg~s^{-1},
	\label{eq:lmin}
	\end{split}
\end{equation} 
where R, M, $\mu$ and P$_s$ are the radius, mass, magnetic moment and spin period of the NS, respectively. 

We use a surface polar magnetic field strength $B=2.6\times 10^{12}$ G \citep{mih91} and $R = 10$ km, for a dipole-like field configuration, $\mu=B\times R^3=2.6\times 10^{30}$ G cm$^3$.
Assuming $M=1.4 M_{\odot}$, we calculate the minimum luminosity below which the propeller effect will occur in SXP 1323 to be $L_\mathrm{X} (\mathrm{min})=7.03\times 10^{30}$ {\lx}. In our analysis, all of the luminosities observed are higher than this critical value, therefore it is highly unlikely that the anti-correlation is the result of the propeller effect.
   
 We can see the PF drops quickly as the luminosity increases up to $\sim$$10^{36}$~{\lx}. This is consistent with \citet{cam01}'s result above a certain critical luminosity of $\sim$$10^{35}$~{\lx} in the X-ray pulsar 4U 0115+63 in our galaxy. \citet{cam01} expressed the source accretion luminosity as two components: the luminosity of the disk extending down to the magnetospheric boundary, $L_{disk}$; and the luminosity released within the magnetosphere $L_{mag}$ by the mass inflow rate that accretes onto the NS surface. They claimed that the pulsed fraction is expected to remain unaltered as long as $L_{mag}$ dominates, while $L_{disk}$ is expected to be un-pulsed, resulting in a decreasing pulsed fraction as its luminosity increases. It explains the PF trend only at the luminosities larger than $\sim$$10^{35}$~{\lx} in Fig. 2 of \citet{cam01}.

   Assuming that there are two X-ray components: the accretion column ($L_\mathrm{col}$) and the accretion disk ($L_\mathrm{disk}$), the luminosity of the accretion column should be relatively stable since it would be locally Eddington, and the luminosity of the disk changes because at high mass accretion rate ($\dot M$) the magnetospheric radius ($R_\mathrm{mag}$) gets smaller and the $L_\mathrm{disk}$ increases.
   
   From the relation between $R_{mag}$ and $\dot M$ (for a given $P_s$ and magnetic field B) and feeding this through a standard Shakura-Sunyaev disk, we have that:
   
   \begin{equation}
  \begin{split}
  T_\mathrm{disk} \propto \left \{  \frac{\dot M}{R^3}[1-(\frac{R_\mathrm{mag}}{R})^{\frac{1}{2}}] \right \}^{\frac{1}{4}}
  	\label{eq:lmin}
	\end{split}
\end{equation} 
   
    \begin{equation}
  \begin{split}
    L_\mathrm{disk} \propto T_\mathrm{disk}^{4}  \centerdot R^{2}\simeq \dot M \centerdot  R^{\frac{5}{4}}
	\label{eq:lmin}
	\end{split}
\end{equation} 

If luminosity from the accretion column $L_\mathrm{col}$ is constant, $R_\mathrm{mag}$ decreases, and $L_\mathrm{disk}$ increases. The predicted PF should change with increasing luminosity due to the un-pulsed disk emission. 

   Our anti-correlation is still at odds with the trend reported for many other pulsars in the literature \citep[e.g.,][]{muk00, coe15}. The possible reason is that the spin period matters, as the pulsars in \citet{muk00} and \citet{coe15} have short spin periods of 66.27s and 5.05s, respectively. It could be that the PF changes of the short period pulsars depend on their luminosities.
  
   
   In the following, we discuss the PF-luminosity anti-correlation in the context of different models for X-ray emission in accreting pulsars.

\subsection{Spherical accretion}
The flow of material toward the pulsar might not take place through an accretion disk but instead via a spherical accretion flow, a natural consequence of wind-fed accretion, as opposed to Roche-lobe overflow. The spherical accretion should be outside the accretion column and would obscure it (unless highly ionized). Also at low luminosity, the magnetospheric radius should be large enough to truncate the accretion flow. This accretion model was applied to black holes by \citet{nob91}. The accretion of gas onto the compact object can be a very efficient way of converting gravitational potential energy into radiation. Traditional spherical accretion is thought of as a good approximation for isolated compact objects.  
\citet{ikh12} has applied the spherical accretion model to HMXBs, especially the long spin period pulsars. \citet{zel69} presented a model to describe the gravitational energy of matter accreted onto a NS and released in a thin layer above the surface. Variations of this idea have also been applied in detailed models \citep[e.g.,][for spherical accretion]{tur94}. The deep layers of the NS atmosphere are heated by the outer layer and produce soft thermal photons \citep{cui98}. The hard X-ray photons are from the polar hot spots, which contribute to the pulsed flux observed. The soft X-ray photons from spherical accretion would mainly contribute to the un-pulsed component of flux. Spherical accretion becomes more prominent as the luminosity and mass accretion rate increases, which leads to a smaller PF. 


\subsection{NS whole surface thermal emission}
Generally, there are two components of the X-ray emission from NSs: thermal emission and non-thermal emission. The non-thermal emission is caused by pulsar radiation in the magnetosphere and its own rotation activity, which is suppressed when accreting. Thermal emission is from the whole surface of a cooling NS and/or from the small hot spots around the magnetic poles on the star surface \citep{bec09}. It is also heated by accretion. The thermal radiation from the entire stellar surface can dominate at soft X-ray energies for middle-aged pulsars ($\sim$100 kyr) and younger pulsars ($\sim$10 kyr). 

If thermal emission is a significant component of the X-rays from SXP 1323 and this component increases, it would represent and increase in un-pulsed flux such that the PF becomes smaller.

\subsection{Change in emission geometry}
\citet{gho79} found $\dot{P}\propto L_\mathrm{X} ^{6/7}$ assuming the effective inertial moment of the NS is constant, so a higher accretion rate and $L_\mathrm{X} $ could cause the observed rapid spin-up rate of this pulsar. The accreted mass interacts with the magnetosphere and the accretion disk extends inward to some equilibrium radial distance above the NS's surface \citep{mal89}. \citet{yan17} has reported this pulsar's average spin-up rate as $6 \pm 3$ millisecond/day based on data from 3 X-ray satellites from 1997-2014. \citet{car17} has presented an even faster spin-up of $\sim$$59.3$ millisecond/day based on recent observations from 2006 to 2016. The higher spin period and mass accretion rate could build up a higher accretion column above the polar caps. As the height of the accretion column increases, scattering of photons off in-falling electrons gets more prominent. This increases the fraction of emission escaping the column to the side, i.e., a fan-beam emerges \citep[e.g.,][]{bec12}. Fan-beam emission becomes dominant, which reduces the eclipse of the accretion column. Furthermore, the contribution of the flux reflected by the NS surface is significant \citep{mus18}. It raises the un-pulsed flux, therefore we see the luminosity increasing and the pulsed fraction decreasing. 

\citet{rom08} used 3D MHD simulations for a star that might be in the stable or unstable regime of accretion. In the unstable regime, matter penetrates into the magnetosphere and is deposited at random places on the surface of the star, which made the pulsations intermittent or with no pulsations. Therefore, the PF is reduced when the overall X-ray flux increases which may be also due to the transition to the unstable accretion regime.

In this scenario, we predict that the slope of the PF versus the luminosity will decrease as the spin periods of the pulsars increase. We will further investigate all of the pulsars in our current library to test this prediction. 

\section{Conclusions}


The anti-correlation between the PF and luminosity in SXP 1323 reveals that different accretion modes are possible. This could be related to the puzzle of the existence of long period pulsars
which are hard to explain \citep{Ikhsanov2014} without invoking non-standard accretion modes (such as spherical accretion). However, the significance of the anti-correlation is not high enough to prove its existence. SXP 1323 is the best example within our sample and more high quality data from the future observations are still needed to check up on the anti-correlation.

\section*{Acknowledgements}
We would like to thank the anonymous referee whose valuable suggestions and comments have significantly improved the quality of the paper.
\bsp	
\label{lastpage}
\end{document}